\documentclass{llncs} 
\usepackage{makeidx}
\usepackage{epsf}
\usepackage{latexsym}
\usepackage{amsmath, amssymb}
\usepackage{epsfig}
\begin{document}
\title{Intermittency and Obsolescence: a Croston Method With Linear Decay}
\author{S. D. Prestwich${}^1$,
S. A. Tarim${}^2$,
R. Rossi${}^3$,
B. Hnich${}^4$\\
{\small \it ${}^1$Insight Centre for Data Analytics, University College Cork, Ireland}\\
{\small \it ${}^2$Institute of Population Studies, Hacettepe University, Ankara, Turkey}\\
{\small \it ${}^3$University of Edinburgh Business School, Edinburgh, UK}\\
{\small \it ${}^4$Computer Engineering Department, Izmir University of Economics, Turkey}
}
\institute{}
\maketitle
\begin{abstract}

Only two Croston-style forecasting methods are currently known for
handling stochastic intermittent demand with possible demand
obsolescence: TSB and HES, both shown to be unbiased.  When an item
becomes obsolescent then TSB's forecasts decay exponentially, while
HES's decay hyperbolically.  We describe a third variant called
Linear-Exponential Smoothing that is also unbiased, decays linearly to
zero in a finite time, is asymptotically the best variant for handling
obsolescence, and performs well in experiments.

\vspace{2mm}

\end{abstract}

\section{Introduction}

Inventory management is of great economic importance to industry, but
forecasting demand for spare parts is difficult because it is {\it
  intermittent\/}: in many time periods the demand is zero.  This type
of demand occurs in several industries, for example in aerospace and
military inventories from which spare parts such as wings or jet
engines are infrequently required.  Various methods have been proposed
for forecasting, some simple and others statistically sophisticated,
but relatively little work has been done on intermittent demand.  Most
work in this area is influenced by that of \cite{Cro}, who first
separated the forecasting of demand size and inter-demand interval.  A
recent review of the literature on intermittent demand can be found in
\cite{SnyEtc}.

Another difficult feature of some inventories is {\it obsolescence\/},
in which an item has no demand at all after a certain time period.
When many thousands of items are being handled automatically, this may
go unnoticed by Croston-style methods, which continue to forecast high
demand forever though no actual demand has occurred.  The authors of
this paper know of an inventory company who were obliged to modify
Croston's method, artificially forcing its forecasts to zero after a
certain number of periods without demand.  This is a pragmatic but
inelegant solution, and obsolescence has been neglected in the
literature.  However, two recent Croston variants have been designed
to tackle it: TSB \cite{TeuEtc} and HES \cite{PreEtc}.

A qualitative difference between TSB and HES is that when obsolescence
occurs TSB's forecasts decay exponentially to zero while those of HES
decay hyperbolically.  Neither generates forecasts that actually reach
zero, though they come arbitrarily close as time proceeds.  In this
paper we describe a new Croston variant whose forecasts decay linearly
to zero in a finite time, a feature we believe will appeal to
practitioners.  We compare it empirically and analytically with other
forecasters and show that it is unbiased, handles obsolescence better
than other methods, and competitive in experiments with intermittent
demand.

The paper is organised as follows.  Section \ref{background} surveys
existing forecasters and presents the new forecaster, Section
\ref{asymp} analyses the handling of obsolescence by forecasters,
Section \ref{experiments} compares them empirically using synthetic
demand data, and Section \ref{conc} concludes the paper.

\section{Forecasting for intermittency and obsolescence} \label{background}

In this section we survey the relevant forecasting methods for
handling intermittency and obsolescence, and introduce our new
forecaster.  We denote the observed demand at (discrete) time $t$ by
$y_t$, a smoothed estimate of $y$ by $\hat{y}_t$, and a forecast by
$f_t$.

{\it Single exponential smoothing\/} (SES) generates estimates
$\hat{y}_t$ of the demand by exponentially weighting previous
observations using the formula
\[
\hat{y}_t = \alpha y_t + (1-\alpha)\hat{y}_{t-1}
\]
where $\alpha \in (0,1)$ is a {\it smoothing parameter\/}.  The
smaller the value of $\alpha$ the less weight is attached to the most
recent observations.  There are many variations on SES and they are
surveyed in \cite{Gar}.  They perform remarkably well, often beating
more sophisticated approaches \cite{FilEtc1}, but SES is known to
perform poorly on {\it stochastic intermittent demand\/}.  In a
standard model of this type of demand, the occurrence of a nonzero
demand is a Bernoulli event occurring at each time period with some
probability.  The magnitude of the demands may follow any of several
distributions.

A well-known method for handling intermittency is {\it Croston's
  method\/} \cite{Cro} which explicitly separates the aspects of
demand size and probability of a demand occurring.  It applies SES to
the demand size $y$ and inter-demand interval $\tau$ independently
(possibly with different smoothing factors), where $\tau=1$ for
non-intermittent demand.  Given smoothed demand $\hat{y}_t$ and
smoothed interval $\hat{\tau}_t$ at time $t$, the forecast is
\[
f_t = \frac{\hat{y}_t}{\hat{\tau}_t}
\]
Both $\hat{y}_t$ and $\hat{\tau}_t$ are updated at each time $t$ for
which $y_t \neq 0$.  According to \cite{Gar} it is hard to conclude
from the various studies that Croston's method is successful, because
the results depend on the data used and on how forecast errors are
measured.  But it is generally regarded as one of the best methods for
intermittent series \cite{GhoFri}, and versions of the method are used
in leading statistical forecasting software packages \cite{TeuEtc}.
We refer to it as CR.

CR was shown by \cite{SynBoy} to be biased on stochastic intermittent
demand, and they corrected the bias by modifying the forecasts:
\[
f_t = \left( 1 - \frac{\beta}{2} \right) \frac{\hat{y}_t}{\hat{\tau}_t}
\]
where $\beta$ is the smoothing factor used for inter-demand intervals,
which may be different to the $\alpha$ smoothing factor used for
demands.\footnote{In \cite{SynBoy} this factor is denoted by $\alpha$
  because it is used to smooth both $\hat{y}$ and $\hat{\tau}$.}  We
refer to this variant as SBA.  It works well for intermittent demand
but is biased for non-intermittent demand, as its forecasts are those
of SES multiplied by $(1- \beta/2)$.  This problem is avoided by
\cite{Syn} who uses a forecast
\[
f_t = \left( 1-\frac{\beta}{2} \right)
\frac{\hat{y}_t}{\hat{\tau}_t- \beta / 2}
\]
This removes the bias on non-intermittent demand but increases the
forecast variance \cite{TeuSan}.  We refer to this variant as SY.

Another modified Croston method is described by \cite{LevSeg}, who
apply SES to the ratio of demand size and inter-demand period when a
nonzero demand occurs:
\[
f_t = \alpha \left(\frac{y_t}{\tau_t}\right) + (1-\alpha) f_{t-1}
\]
However, this turns out to be biased on stochastic intermittent demand
\cite{BoySyn}.

Though these variants successfully handle intermittency, they do not
handle obsolescence well: when obsolescence occurs they continue
forever to forecast a fixed nonzero demand.  The first Croston variant
explicitly designed to handle obsolescence is the TSB method of
\cite{TeuEtc} which updates an estimate of the {\it demand
  probability\/} instead of the inter-demand interval: instead of a
smoothed interval $\hat{\tau}_t$ it uses a smoothed probability
estimate $\hat{p}_t$ where $p_t$ is 1 when demand occurs at time $t$
and 0 otherwise.  Different smoothing factors $\alpha$ and $\beta$ are
used for $\hat{y}_t$ and $\hat{p}_t$ respectively.  $\hat{p}_t$ is
updated every period while $\hat{y}_t$ is only updated when demand
occurs.  The forecast is
\[
f_t=\hat{p}_t\hat{y}_t
\]
This method is unbiased and handles intermittency well.  It also
solves the problem of obsolescence because, like SES but unlike other
Croston variants, when an item becomes obsolescent its forecasts decay
exponentially to zero.

Another Croston variant designed to handle obsolescence is the
Hyperbolic-Exponential Smoothing (HES) method of \cite{PreEtc}.  Like
most Croston variants HES separates demands into demand size $y_t$ and
inter-demand interval $\tau_t$.  Its forecasts are
\[
f_t = \left\{
\begin{array}{l@{\hspace{5mm}}l}
\hat{y}_t / \hat{\tau}_t & \mbox{if $y_t>0$}\\
\hat{y}_t / (\hat{\tau}_t + \beta\tau_t/2) & \mbox{if $y_t=0$}
\end{array}
\right.
\]
Between nonzero demands $\tau$ increases linearly, producing a
hyperbolic decay in the forecasts.  This was justified in
\cite{PreEtc} by a Bayesian argument.

Our new Croston variant is similar in form to HES but uses forecasts
\[
f_t = \left\{
\begin{array}{l@{\hspace{5mm}}l}
\hat{y}_t / \hat{\tau}_t & \mbox{if $y_t>0$}\\
(\hat{y}_t / \hat{\tau}_t) (1- \beta \tau_t / 2\hat{\tau}_t)^+ & \mbox{if $y_t=0$}
\end{array}
\right.
\]
where $x^+$ denotes $\max(0,x)$.  When obsolescence occurs the
forecasts decay linearly to zero at a rate controlled by $\beta$, and
when they reach zero they remain there until further nonzero demands
occur.  The rate at which they decay can be controlled by adjusting
$\beta$.  This feature distinguishes it from all other Croston
variants, which only approach zero asymptotically.  We call this
forecaster Linear-Exponential Smoothing (LES).

We show in Appendix \ref{lesder} that LES is theoretically unbiased on
stochastic intermittent demand, under the assumption that $1-
\beta\tau_t / 2\hat{\tau}_t \ge 0$.  If this assumption does not hold
(which may occur if we set $\beta$ to a high value) then the term will
be replaced by 0, causing a positive bias, but we show empirically in
Section \ref{stationary} that this effect is negligible.

\begin{figure}
\begin{center}
\includegraphics[scale=0.9]{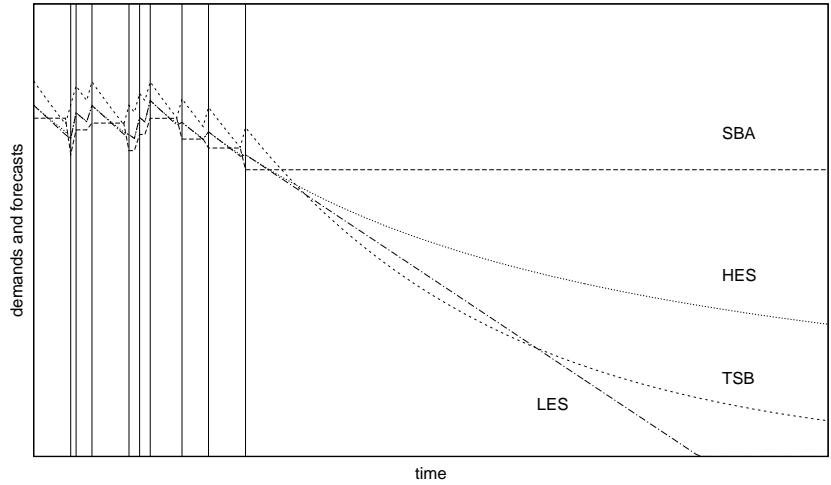}
\end{center}
\caption{SBA, TSB, HES and LES under sudden obsolescence}
\label{pluseffect}
\end{figure}

Pseudocode for LES is shown in Figure \ref{lescode} and a graph
illustrating its behaviour is shown in Figure \ref{pluseffect}.  At
the left of the graph demand is stochastic intermittent (shown as
impulses) with probability 0.25 and fixed size, but then sudden
obsolescence occurs as the probability drops instantaneously to 0.
All forecasters use $\alpha= \beta = 0.1$, except that TSB uses
$\beta=0.02$ because \cite{TeuEtc} recommend a smaller value.  The
graph shows that all four forecasters behave reasonably on stationary
demand, but that when obsolescence occurs SBA (like most Croston
variants) continues indefinitely with a nonzero forecast while TSB,
HES and LES decay in different ways.

\begin{figure}
\begin{tabbing}
aaaaaaaaaaaaaaaaaaaa\=aaa\=aaa\=aaa\=aaa\=aaa\=aaa\=aaa\=\kill
\>$\hat{y} \leftarrow 1$, $\tau \leftarrow 1$, $\hat{\tau} \leftarrow 1$\\
\>at each time period\\
\>\>$y \leftarrow$ current demand\\
\>\>if $y \neq 0$\\
\>\>\>$\hat{y} \leftarrow \alpha y + (1-\alpha) \hat{y}$\\
\>\>\>$\hat{\tau} \leftarrow \beta \tau + (1-\beta) \hat{\tau}$\\
\>\>\>$f \leftarrow \hat{y} / \hat{\tau}$\\
\>\>\>$\tau \leftarrow 1$\\
\>\>else\\
\>\>\>$f \leftarrow (\hat{y} / \hat{\tau})(1 - \beta \tau / 2 \hat{\tau})^+$\\
\>\>\>$\tau \leftarrow \tau+1$
\end{tabbing}
\caption{Pseudocode for LES}
\label{lescode}
\end{figure}

\section{Asymptotic obsolescence error} \label{asymp}

There are now three Croston variants that are explicitly designed to
handle obsolescence: TSB, HES and LES.  They have qualitatively
different behaviour when obsolescence occurs, respectively decaying
exponentially, hyperbolically and linearly.  Each is approximately
unbiased on stochastic intermittent demand, but which best handles
obsolescence?

This is a difficult question because the answer clearly depends on
many factors: the type of demand data, how long we compare forecasters
before and after obsolescence occurs, and which error measures we use
for the comparison.  In Section \ref{experiments} we shall perform
experiments, but in this section we analyse the asymptotic behaviour
of the different forecasters, in an attempt to obtain a definitive
answer.

We shall compute error measures for the forecasters, using times
starting from just after obsolescence occurs at time 0, up to some
large $T \rightarrow \infty$.  We assume the demand to be highly
intermittent, that is $\tau_t$ is typically large, so the user will
choose small $\beta$.  This represents a worst-case scenario in which
an automated inventory control system continues to make forecasts far
from zero for an obsolete item for a long time, because it believes
demand to be highly intermittent based on previous data.  We shall
analyse how the forecasters perform under this scenario.  We ignore
the machine-dependent issue of arithmetic errors causing truncation to
0 as forecasts become small.  All the forecasters are unbiased so we
assume they have the same forecast $f_0$ when obsolescence occurs at
time 0.

A surprising variety of measures have been used in the literature and
in forecasting competitions \cite{MakEtc1,MakEtc2,MakHib}.  There is
no consensus on which is best so it is generally recommended to use
several.  We shall consider all measures listed in the surveys of
\cite{GooHyn,HynKoe} and the article \cite{WalSeg}.

The {\it scale-dependent measures\/} are based on the mean error
$e_t=y_t- \hat{y}_t$ or mean square error $e^2_t$, and include Mean
Error, Mean Square Error, Root Mean Square Error, Mean Absolute Error
and Median Absolute Error.  As $T \rightarrow \infty$ all these tend
to zero so they cannot be used for an asymptotic comparison.

The {\it percentage errors\/} are based on the quantities
$p_t=100e_t/y_t$ and include Mean Absolute Percentage Error, Median
Absolute Percentage Error, Root Mean Square Percentage Error, Root
Median Square Percentage Error, Symmetric Mean Absolute Percentage
Error, and Symmetric Median Absolute Percentage Error.  As $y_t=0$ for
all $t>0$ these are undefined for almost all times.

The {\it relative error-based measures\/} are based on the quantities
$r_t = e_t / e^*_t$ where $e^*_t$ is the error from a baseline
forecaster, and include Mean Relative Absolute Error, Median Relative
Absolute Error, and Geometric Mean Relative Absolute Error.  The
baseline forecaster is usually the {\it random walk\/} (or {\it naive
  method\/}) which simply forecasts that the next demand will be
identical to the current demand.  For almost all times $e^*_t=0$ so
these measures are undefined.  We could use another baseline but we
would still have the problem that the mean and median $e_t$ are zero,
so these cannot be used for a comparison.

The {\it relative measures\/} are mainly defined as the ratio of (i)
an error measure, and (ii) the same measure applied to a baseline
forecaster.  These include Relative Mean Absolute Error, Relative Mean
Squared Error, and Relative Root Mean Squared Error (for example the
U2 statistic).  The baseline forecaster is again usually the random
walk.  Both measures tend to zero as $T \rightarrow \infty$ so these
cannot be used for our comparison.  A different form of relative
measure is Percent Better, which computes the percentage of times a
forecaster has smaller absolute error $|e_t|$ than a baseline
forecaster, again usually random walk.  Random walk has asymptotically
perfect performance so Percent Better cannot be used for our
comparison.  A related measure is Percent Best in which no baseline
forecaster is used: instead it computes the percentage of times each
forecaster being tested has smaller absolute error than the others.
We shall use this measure below.

The {\it scaled errors\/} include MAD/Mean Ratio \cite{KolSch} and
Mean Absolute Scaled Error \cite{HynKoe}.  The former cannot be used
for our comparison because the denominator (the mean error) tends to
zero, while the latter cannot be used because it is proportional to
$e_t$ which tends to zero.

There are also three recent measures designed for intermittent demand
\cite{WalSeg}.  (i) Cumulative Forecast Error is defined as the sum of
all errors over the time periods under consideration.  Not taking
averages means that errors do not become vanishingly small, so this
measure gives meaningful results.  We shall use it and also the
related Cumulative Squared Error (which was not mentioned in
\cite{WalSeg}): a motivation for using squared errors is that they
penalise outliers more severely than absolute errors, giving a
different perspective.  (ii) Number of Shortages at time $t$ is
defined as the number of periods in which the Cumulative Forecast
Error is strictly positive and demand is nonzero.  In our scenario
demand is always zero after obsolescence occurs so this is not
meaningful.  (iii) Periods in Stock at time $t$ is defined as
\[
\sum_{i=1}^t (\hat{y}_i - y_i)(t+1-i)
\]
In our scenario $y_i=0$ for all $i>0$ so this reduces to
\[
f_0 \lim_{t \rightarrow \infty} \sum_{i=1}^t \hat{y}_i (t+1-i)
\]
But as $t \rightarrow \infty$ the term $\hat{y}_1t \rightarrow \infty$
and all other terms are positive, so this measure is also not
meaningful here.

Thus Percent Best (PBt), Cumulative Forecast Error (CFE) and
Cumulative Squared Error (CSE) are the only error measures we know of
that can be used for our comparison.  The results of the comparison
are shown in Table \ref{asympres} and the derivations are given in
Appendix \ref{errdev}.  First we consider the CFE results.  HES is
worst with infinite error.  TSB appears to beat LES but only if they
use the same smoothing factor $\beta$, and it is recommended by
\cite{TeuEtc} to use a smaller $\beta$ for TSB.  In the absence of an
exact known relationship between $\beta$ and $\hat{\tau}_t$ (we know
only that they are inversely related in some sense) the two results
are incomparable, so neither TSB nor LES can be shown to dominate the
other.  Next we consider the CSE results.  TSB is incomparable with
HES and LES, but LES is 3 times better than HES.  Finally, we consider
the PBt results.  Here LES beats both HES and TSB.

\begin{table}
\begin{center}
\begin{tabular}{cccc}
\hline
& CFE & CSE & PBt\\
\hline
TSB & $f_0 / \beta$
& $f_0^2 / \beta^2$ & 0\%\\
HES & $\infty$ & $2 f_0^2\hat{\tau}_0 / \beta$ & 0\%\\
LES & $f_0\hat{\tau}_0 / \beta$ & $2f_0^2\hat{\tau}_0 / 3\beta$ & 100\%\\
\hline
\end{tabular}
\end{center}
\caption{Asymptotic obsolescence errors}
\label{asympres}
\end{table}

In summary, LES is not beaten by TSB or HES under CFE or CSE, but
beats both under PBt, so we rank LES as the best variant for handling
obsolescence.  TSB beats HES under CFE, draws with it under PBt and is
incomparable under CSE, so we rank it second best.  HES ranks third
best.

\section{Experiments} \label{experiments}

In this section we test the accuracy of LES using synthetic demand
data, to verify that it performs well empirically as well as
theoretically.  All experiments are based on those of \cite{TeuEtc}.

\subsection{Stationary demand} \label{stationary}

First we compare LES with TSB and HES on {\it stationary\/} stochastic
intermittent demand (no obsolescence).  Teunter {\it et al.\/} compare
several forecasters on demand that is nonzero with probability $p_0$
where $p_0$ is either 0.2 or 0.5, and whose size is logarithmically
distributed.  Geometrically distributed intervals are a discrete
version of Poisson intervals, and the combination of Poisson intervals
and logarithmic demand sizes yields a negative binomial distribution,
for which there is theoretical and empirical evidence \cite{SynEtc3}.
The logarithmic distribution is characterised by a parameter $\ell \in
(0,1)$ and is discrete with $\Pr[X=k]= - \ell^k/k \log(1- \ell)$ for
$k \ge 1$.  Teunter {\it et al.\/} use two values: $\ell=0.001$ to
simulate low demand and $\ell=0.9$ to simulate lumpy demand.  They use
several combinations of smoothing factors: $\alpha$ values 0.1, 0.2
and 0.3, and $\beta$ values 0.01, 0.02, 0.03, 0.04, 0.05, 0.1, 0.2,
0.3.  We use 1000 runs for each forecaster, with 1000 time periods
used for initialisation and a further 1000 periods for evaluation.  We
use three error measures: Mean Error (ME) to measure bias, Mean
Absolute Error (MAE) and Root Mean Square Error (RMSE).

The results are shown in Tables \ref{sta1}--\ref{sta4}.  We compare
the forecasters by considering best results as $\alpha$ and $\beta$
are varied.  TSB and HES have lowest bias (ME), while HES and LES have
lowest deviation (MAE and RMSE).  The ME results also show that LES
has low bias (though not the lowest) despite the fact that, as noted
in Section \ref{background}, it will not be unbiased if the term
$1-\beta \tau_t/2 \hat{\tau}_t$ becomes negative.

\subsection{Decreasing demand} \label{lindec}

In this experiment demand sizes again follow the logarithmic
distribution, but the probability of a nonzero demand decreases
linearly from $p_0$ in the first period to 0 during the last period.
Demand sizes are again logarithmically distributed.  As pointed out by
Teunter {\it et al.\/}, none of the forecasters use trending to model
the decreasing demand so all are positively biased.  The results are
shown in Appendix \ref{expres}, Tables \ref{dec1}--\ref{dec4}.  Under
ME, MAE and RMSE, TSB ranks first, LES second and HES third.

\subsection{Sudden obsolescence} \label{sudden}

This experiment is the same as that of Section \ref{lindec} except
that the demand probability is reduced instantly to 0 after half the
time periods.  Demand sizes are again logarithmically distributed.
The results are shown in Tables \ref{obs1}--\ref{obs4}.  LES wins
under ME and MAE, and TSB under RMSE.

\subsection{Summary}

Table \ref{winners} summarises the winning forecaster under each
combination of demand pattern and error measure.  The results
illustrate the lack of clarity caused by mutiple factors in the
experiments, as no clear pattern emerges.  All three perform well on
stationary demand, TSB is the clear winner under decreasing demand,
and LES wins more often under sudden obsolescence.  However, we can
conclude that LES is highly competitive under three error measures.

\begin{table}
\begin{center}
\begin{tabular}{lccc}
\hline
demand     & ME & MAE & RMSE\\
\hline
stationary & TSB+HES & HES+LES & HES+LES\\
decreasing & TSB & TSB & TSB\\
sudden     & LES & LES & TSB\\
\hline
\end{tabular}
\end{center}
\caption{Winning forecasters under different conditions}
\label{winners}
\end{table}

\section{Conclusion} \label{conc}

We described a new Croston variant called LES for handling
obsolescence, shown to be unbiased on stochastic intermittent demand.
LES has a feature that we consider to be an advantage over the two
other variants TSB and HES designed to handle obsolescence: when
obsolescence occurs its forecasts decay to zero in a finite time.
This also occurs when a non-intermittent item becomes obsolescent, so
LES may be a useful alternative to SES for non-intermittent as well as
intermittent demand.

We proposed a form of asymptotic analysis to compare how well
forecasters handle obsolescence, based on a worst-case scenario in
which a highly-intermittent item becomes obsolescent and forecasts
continue forever.  Our analysis ranks LES as the best variant,
followed by TSB then HES.

Finally, we performed experiments using synthetic demand data, and
found LES to be highly competitive compared to TSB and HES.  TSB has
previously been shown to have lower bias and deviation than other
Croston variants \cite{TeuEtc} so LES will also compare well against
these forecasters.

\section*{Acknowledgment}

This publication has emanated from research supported in part by a
research grant from Science Foundation Ireland (SFI) under Grant
Number SFI/12/RC/2289.  S. Armagan Tarim is supported by the
Scientific and Technological Research Council of Turkey (TUBITAK)
under Grant No. 110M500.  R. Rossi is supported by the University of
Edinburgh CHSS Challenge Investment Fund and by the European
Community's Seventh Framework Programme (FP7) under grant no 244994
(project VEG-i-TRADE).

\bibliographystyle{plain}

\appendix

\section{Derivation of the forecaster} \label{lesder}

This derivation follows a similar pattern to that of HES
\cite{PreEtc}.  The LES forecaster uses a forecast of the form
\[
f_t = \left\{
\begin{array}{l@{\hspace{5mm}}l}
\hat{y}_t / \hat{\tau}_t & \mbox{if $y_t>0$}\\
(\hat{y}_t / \hat{\tau}_t) (1- k \tau_t)^+ & \mbox{if $y_t=0$}
\end{array}
\right.
\]
for some fixed value $k$, and we choose $k$ to make LES unbiased on
stochastic intermittent demand.  First we derive the expectation
$\mathbb{E}[f_t]$.  Consider the demand sequence as a sequence of {\it
  substrings\/}, each starting with a nonzero demand: for example the
sequence $(5,0,0,1,0,0,0,3,0)$ has substrings $(5,0,0)$, $(1,0,0,0)$
and $(3,0)$.  Within a substring $\hat{y}_t$ and $\hat{\tau}_t$ remain
constant, and if an item has not become obsolescent and $k$ is
sufficiently small then $1- k \tau_t>0$, so LES has expected forecast
\[
\mathbb{E}\left[ \left( \frac{\hat{y}_t}{\hat{\tau}_t} \right)
(1 - k \tau_t) \right] =
\left( \frac{\hat{y}_t}{\hat{\tau}_t} \right)
\left( 1 - k \mathbb{E} \left[\tau_t \right] \right)
\]
For stochastic intermittent demand the inter-demand interval is a
random variable with geometric distribution and mean $1/p$.  We
estimate $p \approx 1/ \hat{\tau}_t$ so $\mathbb{E} \left[\tau_t
  \right] \approx \hat{\tau}_t$ and the expected forecast over the
string is
\[
\left( \frac{\hat{y}_t}{\hat{\tau}_t} \right)
\left( 1 - k \hat{\tau}_t \right)
\]
This coincides with SES on non-intermittent demand, so LES is unbiased
on non-intermittent demand whatever the value of $k$.  To make it
unbiased on stochastic intermittent demand we choose $k$ so that it
has the same expected forecast as SBA's fixed forecast over each
string, which is
\[
\left( \frac{\hat{y}_t}{\hat{\tau}_t} \right)
\left(1 - \frac{\beta}{2} \right)
\]
So $k = \beta / 2 \hat{\tau}_t$ and the forecast when $y_t=0$ is
\[
f_t = \left( \frac{\hat{y}_t}{\hat{\tau}_t} \right)
\left(1 - \frac{\beta \tau_t}{2 \hat{\tau}_t} \right)^+
\]
Moreover, LES updates $\hat{y}_t$ and $\hat{\tau}_t$ in exactly the
same way as SBA at the start of each substring, therefore it has the
same expected forecast as SBA over the entire demand sequence.  Thus
by \cite{SynBoy} it is unbiased on stochastic intermittent demand.

\section{Derivation of asymptotic errors} \label{errdev}

In this Appendix we derive asymptotic obsolescence errors for the
three forecasters.

\subsection{Cumulative forecast error} \label{cfecomp}

The CFE is the sum of all errors for $t \ge 0$, used for example in
\cite{WalSeg}.  In our scenario all forecasts are positive and all
demands are zero, so the CFE coincides with the Cumulative {\it
  Absolute\/} Error.  TSB's CFE is
\[
f_0[1 + (1 - \beta) + (1 - \beta)^2 + \ldots] =
\frac{f_0}{\beta}
\]
HES's CFE is
\[
f_0 \sum_{t=0}^{\infty} \frac{1}{1+t \beta / 2 \hat{\tau}_0}
\]
This is a special case of the general harmonic series which diverges
to $\infty$.  LES's CFE is
\[
f_0 + f_0 \left(1 - \frac{\beta}{2 \hat{\tau}_0}\right) +
f_0 \left(1 - \frac{2 \beta}{2\hat{\tau}_0} \right) + \ldots +
f_0 \left(\frac{\beta}{2 \hat{\tau}_0}\right)
\]
Under the simplifying assumption that $2 \hat{\tau}_0 / \beta$ is an
integer $\ell$ the series contains $\ell$ terms so the CFE is
\[
\frac{f_0}{\ell} \left[ \ell + \left( \ell - 1 \right) +
\left( \ell - 2 \right) + \ldots + 1 \right] =
\frac{f_0(\ell+1)}{2} \approx \frac{f_0\ell}{2} = \frac{f_0\hat{\tau}_0}{\beta}
\]

\subsection{Cumulative squared error} \label{csecomp}

The CSE is the sum of all squared errors.  TSB's CSE is
\[
f_0^2 [1 + (1 - \beta)^2 + (1 - \beta)^4 + \ldots] =
\left( \frac{f_0}{\beta} \right)^2
\]
HES's CSE is
\[
f_0^2 \sum_{t=0}^{\infty} \frac{1}{(1+t\beta/2\hat{\tau}_0)^2}
\]
To evaluate this summation we use the digamma function.  It is known
that
\[
\psi^{(1)}(z) = \sum_{i=0}^{\infty} \frac{1}{(z+i)^2}
\]
where $\psi^{(1)}$ is the first derivative of the digamma function.
Replacing $z$ by $1/x$:
\[
\psi^{(1)}(1/x) = \sum_{i=0}^{\infty} \frac{1}{(1/x+i)^2}
= x^2 \sum_{i=0}^{\infty} \frac{1}{(1+ix)^2}
\]
so
\[
\sum_{i=0}^{\infty} \frac{1}{(1+ix)^2} = \frac{\psi^{(1)}(1/x)}{x^2}
\]
Using an asymptotic expansion for large $z$:
\[
\psi^{(1)}(z) = \sum_{i=0}^{\infty} \frac{B_i}{z^{i+1}}
\]
where the $B_i$ are Bernoulli numbers.  Taking a first term
approximation we get $B_0/z=B_0x=x$.  Therefore
\[
\frac{\psi^{(1)}(1/x)}{x^2} \approx \frac{1}{x}
\]
Substituting $\beta/2\hat{\tau}_0$ for $x$, HES's CSE is $2 f_0^2
\hat{\tau}_0/ \beta$.  LES's CSE is
\[
\left( \frac{f_0}{\ell} \right)^2
\left[ \ell^2 + \left( \ell - 1
\right)^2 + \left( \ell - 2 \right)^2 + \ldots + 1 \right] =
\left( \frac{f_0}{\ell} \right)^2
\left( \frac{\ell^3}{3} + \frac{\ell^2}{2} + \frac{\ell}{6} \right)
\]
Recall that $\ell=2\hat{\tau} / \beta$, which for highly intermittent
demand is a large number, so we ignore the $\ell^2$ and $\ell$ terms
to get $2f_0^2\hat{\tau}_0/3 \beta$.

\subsection{Percent best} \label{pbtcomp}

To compute Percent Best (PBt) we take a collection of forecasting
methods and count the percentage of times at which each gives the
smallest error.  PBt is popular because it is scale-free and easy to
understand.  Furthermore, in practice only one forecaster will be
chosen so PBt resembles a real-world choice \cite{SynBoy}.  Comparing
the three forecasters in this way, LES has a PBt of 100\% while the
others have a PBt of 0\%.  This is because for almost all times both
the demand and the LES forecast are zero while the TSB and HES
forecasts are nonzero.

\section{Experimental results} \label{expres}

In this Appendix we present tables of results for the three
forecasters using synthetic demand data.

\begin{table}
\begin{tabular}{rrrrrrrrrrr}
\hline
&& \multicolumn{3}{c}{TSB} & \multicolumn{3}{c}{HES} & \multicolumn{3}{c}{LES}\\
$\alpha$ & $\beta$ & ME & MAE & RMSE & ME & MAE & RMSE & ME & MAE & RMSE\\
\hline
0.10 & 0.01 & -0.0196 & 2.3253 & 3.8432 & -0.0445 & 2.3112 & 3.8405 & -0.0446 & 2.3112 & 3.8405\\
0.10 & 0.02 & -0.0056 & 2.3364 & 3.8466 & -0.0193 & 2.3255 & 3.8431 & -0.0194 & 2.3254 & 3.8431\\
0.10 & 0.03 & -0.0033 & 2.3421 & 3.8497 & -0.0099 & 2.3321 & 3.8448 & -0.0102 & 2.3320 & 3.8448\\
0.10 & 0.04 & -0.0037 & 2.3465 & 3.8531 & -0.0062 & 2.3361 & 3.8462 & -0.0067 & 2.3359 & 3.8462\\
0.10 & 0.05 & -0.0048 & 2.3503 & 3.8566 & -0.0050 & 2.3390 & 3.8476 & -0.0057 & 2.3388 & 3.8477\\
0.10 & 0.10 & -0.0088 & 2.3667 & 3.8750 & -0.0073 & 2.3497 & 3.8558 & -0.0102 & 2.3486 & 3.8560\\
0.10 & 0.20 & -0.0100 & 2.3983 & 3.9111 & -0.0098 & 2.3662 & 3.8747 & -0.0217 & 2.3613 & 3.8755\\
0.10 & 0.30 & -0.0097 & 2.4317 & 3.9470 & -0.0049 & 2.3817 & 3.8947 & -0.0335 & 2.3719 & 3.8966\\
\hline
0.20 & 0.01 & -0.0065 & 2.3609 & 3.9020 & -0.0318 & 2.3458 & 3.8969 & -0.0318 & 2.3458 & 3.8969\\
0.20 & 0.02 & 0.0082 & 2.3728 & 3.9077 & -0.0060 & 2.3612 & 3.9022 & -0.0061 & 2.3611 & 3.9022\\
0.20 & 0.03 & 0.0106 & 2.3781 & 3.9116 & 0.0039 & 2.3684 & 3.9054 & 0.0037 & 2.3683 & 3.9054\\
0.20 & 0.04 & 0.0099 & 2.3817 & 3.9153 & 0.0078 & 2.3724 & 3.9077 & 0.0074 & 2.3723 & 3.9077\\
0.20 & 0.05 & 0.0083 & 2.3847 & 3.9188 & 0.0092 & 2.3751 & 3.9096 & 0.0085 & 2.3748 & 3.9096\\
0.20 & 0.10 & 0.0018 & 2.3977 & 3.9359 & 0.0057 & 2.3824 & 3.9180 & 0.0028 & 2.3812 & 3.9180\\
0.20 & 0.20 & -0.0018 & 2.4278 & 3.9706 & -0.0002 & 2.3950 & 3.9342 & -0.0122 & 2.3903 & 3.9343\\
0.20 & 0.30 & -0.0023 & 2.4631 & 4.0081 & 0.0030 & 2.4094 & 3.9522 & -0.0259 & 2.3994 & 3.9526\\
\hline
0.30 & 0.01 & 0.0076 & 2.4038 & 3.9683 & -0.0178 & 2.3887 & 3.9608 & -0.0179 & 2.3886 & 3.9608\\
0.30 & 0.02 & 0.0224 & 2.4157 & 3.9762 & 0.0082 & 2.4042 & 3.9690 & 0.0081 & 2.4042 & 3.9690\\
0.30 & 0.03 & 0.0247 & 2.4207 & 3.9810 & 0.0183 & 2.4116 & 3.9737 & 0.0181 & 2.4114 & 3.9736\\
0.30 & 0.04 & 0.0237 & 2.4239 & 3.9850 & 0.0224 & 2.4154 & 3.9768 & 0.0219 & 2.4152 & 3.9768\\
0.30 & 0.05 & 0.0218 & 2.4265 & 3.9887 & 0.0237 & 2.4178 & 3.9792 & 0.0230 & 2.4175 & 3.9792\\
0.30 & 0.10 & 0.0133 & 2.4371 & 4.0051 & 0.0195 & 2.4239 & 3.9883 & 0.0166 & 2.4226 & 3.9881\\
0.30 & 0.20 & 0.0068 & 2.4662 & 4.0383 & 0.0113 & 2.4321 & 4.0028 & -0.0009 & 2.4273 & 4.0021\\
0.30 & 0.30 & 0.0046 & 2.4998 & 4.0768 & 0.0126 & 2.4439 & 4.0190 & -0.0168 & 2.4326 & 4.0175\\
\hline
\end{tabular}
\caption{Stationary stochastic intermittent demand with $\ell=0.9,p_0=0.5$}
\label{sta1}
\end{table}

\begin{table}
\begin{tabular}{rrrrrrrrrrr}
\hline
&& \multicolumn{3}{c}{TSB} & \multicolumn{3}{c}{HES} & \multicolumn{3}{c}{LES}\\
$\alpha$ & $\beta$ & ME & MAE & RMSE & ME & MAE & RMSE & ME & MAE & RMSE\\
\hline
0.10 & 0.01 & -0.0131 & 1.2051 & 2.5216 & 0.0021 & 1.2111 & 2.5190 & 0.0020 & 1.2111 & 2.5190\\
0.10 & 0.02 & -0.0068 & 1.2124 & 2.5248 & -0.0238 & 1.1960 & 2.5191 & -0.0239 & 1.1960 & 2.5191\\
0.10 & 0.03 & -0.0049 & 1.2163 & 2.5276 & -0.0199 & 1.1991 & 2.5199 & -0.0201 & 1.1990 & 2.5199\\
0.10 & 0.04 & -0.0043 & 1.2190 & 2.5303 & -0.0161 & 1.2023 & 2.5206 & -0.0164 & 1.2021 & 2.5206\\
0.10 & 0.05 & -0.0040 & 1.2212 & 2.5329 & -0.0133 & 1.2048 & 2.5213 & -0.0139 & 1.2045 & 2.5213\\
0.10 & 0.10 & -0.0029 & 1.2300 & 2.5459 & -0.0072 & 1.2118 & 2.5241 & -0.0098 & 1.2103 & 2.5241\\
0.10 & 0.20 & -0.0012 & 1.2423 & 2.5740 & -0.0044 & 1.2186 & 2.5289 & -0.0166 & 1.2115 & 2.5290\\
0.10 & 0.30 & -0.0006 & 1.2557 & 2.6055 & -0.0018 & 1.2238 & 2.5336 & -0.0317 & 1.2064 & 2.5350\\
\hline
0.20 & 0.01 & -0.0064 & 1.2167 & 2.5363 & 0.0103 & 1.2251 & 2.5347 & 0.0103 & 1.2251 & 2.5347\\
0.20 & 0.02 & -0.0011 & 1.2231 & 2.5394 & -0.0161 & 1.2086 & 2.5339 & -0.0162 & 1.2085 & 2.5339\\
0.20 & 0.03 & 0.0001 & 1.2265 & 2.5421 & -0.0125 & 1.2115 & 2.5348 & -0.0127 & 1.2114 & 2.5348\\
0.20 & 0.04 & 0.0001 & 1.2288 & 2.5447 & -0.0088 & 1.2145 & 2.5356 & -0.0092 & 1.2143 & 2.5356\\
0.20 & 0.05 & -0.0001 & 1.2306 & 2.5472 & -0.0062 & 1.2167 & 2.5363 & -0.0068 & 1.2164 & 2.5362\\
0.20 & 0.10 & -0.0005 & 1.2382 & 2.5607 & -0.0009 & 1.2230 & 2.5391 & -0.0037 & 1.2214 & 2.5389\\
0.20 & 0.20 & -0.0003 & 1.2509 & 2.5908 & 0.0002 & 1.2277 & 2.5430 & -0.0121 & 1.2204 & 2.5427\\
0.20 & 0.30 & -0.0003 & 1.2635 & 2.6240 & 0.0018 & 1.2324 & 2.5473 & -0.0279 & 1.2149 & 2.5482\\
\hline
0.30 & 0.01 & 0.0029 & 1.2316 & 2.5530 & 0.0204 & 1.2411 & 2.5519 & 0.0204 & 1.2411 & 2.5519\\
0.30 & 0.02 & 0.0076 & 1.2371 & 2.5563 & -0.0063 & 1.2237 & 2.5503 & -0.0064 & 1.2236 & 2.5502\\
0.30 & 0.03 & 0.0082 & 1.2397 & 2.5591 & -0.0028 & 1.2265 & 2.5514 & -0.0030 & 1.2264 & 2.5514\\
0.30 & 0.04 & 0.0077 & 1.2415 & 2.5616 & 0.0008 & 1.2295 & 2.5524 & 0.0004 & 1.2292 & 2.5523\\
0.30 & 0.05 & 0.0070 & 1.2429 & 2.5641 & 0.0033 & 1.2316 & 2.5533 & 0.0027 & 1.2312 & 2.5532\\
0.30 & 0.10 & 0.0048 & 1.2494 & 2.5784 & 0.0081 & 1.2368 & 2.5565 & 0.0053 & 1.2350 & 2.5561\\
0.30 & 0.20 & 0.0031 & 1.2624 & 2.6112 & 0.0080 & 1.2396 & 2.5597 & -0.0046 & 1.2322 & 2.5588\\
0.30 & 0.30 & 0.0020 & 1.2737 & 2.6468 & 0.0084 & 1.2431 & 2.5637 & -0.0211 & 1.2257 & 2.5638\\
\hline
\end{tabular}
\caption{Stationary stochastic intermittent demand with $\ell=0.9,p_0=0.2$}
\label{sta2}
\end{table}

\begin{table}
\begin{tabular}{rrrrrrrrrrr}
\hline
&& \multicolumn{3}{c}{TSB} & \multicolumn{3}{c}{HES} & \multicolumn{3}{c}{LES}\\
$\alpha$ & $\beta$ & ME & MAE & RMSE & ME & MAE & RMSE & ME & MAE & RMSE\\
\hline
0.10 & 0.01 & -0.0062 & 0.4987 & 0.5006 & -0.0121 & 0.4987 & 0.4999 & -0.0121 & 0.4987 & 0.4999\\
0.10 & 0.02 & -0.0028 & 0.4993 & 0.5021 & -0.0063 & 0.4986 & 0.5005 & -0.0063 & 0.4986 & 0.5005\\
0.10 & 0.03 & -0.0018 & 0.4999 & 0.5038 & -0.0041 & 0.4988 & 0.5012 & -0.0042 & 0.4988 & 0.5012\\
0.10 & 0.04 & -0.0014 & 0.5006 & 0.5054 & -0.0031 & 0.4990 & 0.5020 & -0.0032 & 0.4991 & 0.5020\\
0.10 & 0.05 & -0.0012 & 0.5012 & 0.5070 & -0.0025 & 0.4993 & 0.5028 & -0.0027 & 0.4994 & 0.5028\\
0.10 & 0.10 & -0.0009 & 0.5042 & 0.5152 & -0.0017 & 0.5008 & 0.5066 & -0.0024 & 0.5009 & 0.5068\\
0.10 & 0.20 & -0.0007 & 0.5089 & 0.5317 & -0.0009 & 0.5033 & 0.5145 & -0.0040 & 0.5039 & 0.5151\\
0.10 & 0.30 & -0.0006 & 0.5115 & 0.5485 & 0.0008 & 0.5055 & 0.5225 & -0.0069 & 0.5069 & 0.5240\\
\hline
0.20 & 0.01 & -0.0062 & 0.4987 & 0.5006 & -0.0121 & 0.4987 & 0.4999 & -0.0121 & 0.4987 & 0.4999\\
0.20 & 0.02 & -0.0028 & 0.4993 & 0.5021 & -0.0063 & 0.4986 & 0.5005 & -0.0063 & 0.4986 & 0.5005\\
0.20 & 0.03 & -0.0018 & 0.4999 & 0.5038 & -0.0041 & 0.4988 & 0.5012 & -0.0042 & 0.4988 & 0.5012\\
0.20 & 0.04 & -0.0014 & 0.5006 & 0.5054 & -0.0031 & 0.4990 & 0.5020 & -0.0032 & 0.4991 & 0.5020\\
0.20 & 0.05 & -0.0012 & 0.5012 & 0.5070 & -0.0025 & 0.4993 & 0.5028 & -0.0027 & 0.4994 & 0.5028\\
0.20 & 0.10 & -0.0009 & 0.5042 & 0.5152 & -0.0017 & 0.5008 & 0.5066 & -0.0024 & 0.5009 & 0.5068\\
0.20 & 0.20 & -0.0007 & 0.5089 & 0.5317 & -0.0009 & 0.5033 & 0.5145 & -0.0040 & 0.5039 & 0.5151\\
0.20 & 0.30 & -0.0006 & 0.5115 & 0.5485 & 0.0008 & 0.5055 & 0.5225 & -0.0069 & 0.5069 & 0.5240\\
\hline
0.30 & 0.01 & -0.0062 & 0.4987 & 0.5006 & -0.0121 & 0.4987 & 0.4999 & -0.0121 & 0.4987 & 0.4999\\
0.30 & 0.02 & -0.0028 & 0.4993 & 0.5021 & -0.0063 & 0.4986 & 0.5005 & -0.0063 & 0.4986 & 0.5005\\
0.30 & 0.03 & -0.0018 & 0.4999 & 0.5038 & -0.0041 & 0.4988 & 0.5012 & -0.0042 & 0.4988 & 0.5012\\
0.30 & 0.04 & -0.0014 & 0.5006 & 0.5054 & -0.0031 & 0.4990 & 0.5020 & -0.0032 & 0.4991 & 0.5020\\
0.30 & 0.05 & -0.0012 & 0.5012 & 0.5070 & -0.0025 & 0.4993 & 0.5028 & -0.0027 & 0.4994 & 0.5028\\
0.30 & 0.10 & -0.0009 & 0.5042 & 0.5152 & -0.0017 & 0.5008 & 0.5066 & -0.0024 & 0.5009 & 0.5068\\
0.30 & 0.20 & -0.0007 & 0.5089 & 0.5317 & -0.0009 & 0.5033 & 0.5145 & -0.0040 & 0.5039 & 0.5151\\
0.30 & 0.30 & -0.0006 & 0.5115 & 0.5485 & 0.0008 & 0.5055 & 0.5225 & -0.0069 & 0.5069 & 0.5240\\
\hline
\end{tabular}
\caption{Stationary stochastic intermittent demand with $\ell=0.001,p_0=0.5$}
\label{sta3}
\end{table}

\begin{table}
\begin{tabular}{rrrrrrrrrrr}
\hline
&& \multicolumn{3}{c}{TSB} & \multicolumn{3}{c}{HES} & \multicolumn{3}{c}{LES}\\
$\alpha$ & $\beta$ & ME & MAE & RMSE & ME & MAE & RMSE & ME & MAE & RMSE\\
\hline
0.10 & 0.01 & -0.0038 & 0.3318 & 0.4097 & -0.0002 & 0.3331 & 0.4084 & -0.0002 & 0.3331 & 0.4084\\
0.10 & 0.02 & -0.0018 & 0.3338 & 0.4113 & -0.0071 & 0.3292 & 0.4087 & -0.0071 & 0.3292 & 0.4087\\
0.10 & 0.03 & -0.0013 & 0.3347 & 0.4127 & -0.0059 & 0.3301 & 0.4090 & -0.0060 & 0.3301 & 0.4090\\
0.10 & 0.04 & -0.0010 & 0.3353 & 0.4140 & -0.0048 & 0.3310 & 0.4093 & -0.0049 & 0.3309 & 0.4093\\
0.10 & 0.05 & -0.0009 & 0.3356 & 0.4152 & -0.0039 & 0.3316 & 0.4097 & -0.0041 & 0.3316 & 0.4097\\
0.10 & 0.10 & -0.0004 & 0.3368 & 0.4212 & -0.0021 & 0.3335 & 0.4112 & -0.0028 & 0.3331 & 0.4112\\
0.10 & 0.20 & 0.0004 & 0.3383 & 0.4334 & -0.0012 & 0.3349 & 0.4140 & -0.0044 & 0.3331 & 0.4141\\
0.10 & 0.30 & 0.0006 & 0.3397 & 0.4468 & -0.0003 & 0.3357 & 0.4165 & -0.0085 & 0.3311 & 0.4172\\
\hline
0.20 & 0.01 & -0.0038 & 0.3318 & 0.4097 & -0.0002 & 0.3331 & 0.4084 & -0.0002 & 0.3331 & 0.4084\\
0.20 & 0.02 & -0.0018 & 0.3338 & 0.4113 & -0.0071 & 0.3292 & 0.4087 & -0.0071 & 0.3292 & 0.4087\\
0.20 & 0.03 & -0.0013 & 0.3347 & 0.4127 & -0.0059 & 0.3301 & 0.4090 & -0.0060 & 0.3301 & 0.4090\\
0.20 & 0.04 & -0.0010 & 0.3353 & 0.4140 & -0.0048 & 0.3310 & 0.4093 & -0.0049 & 0.3309 & 0.4093\\
0.20 & 0.05 & -0.0009 & 0.3356 & 0.4152 & -0.0039 & 0.3316 & 0.4097 & -0.0041 & 0.3316 & 0.4097\\
0.20 & 0.10 & -0.0004 & 0.3368 & 0.4212 & -0.0021 & 0.3335 & 0.4112 & -0.0028 & 0.3331 & 0.4112\\
0.20 & 0.20 & 0.0004 & 0.3383 & 0.4334 & -0.0012 & 0.3349 & 0.4140 & -0.0044 & 0.3331 & 0.4141\\
0.20 & 0.30 & 0.0006 & 0.3397 & 0.4468 & -0.0003 & 0.3357 & 0.4165 & -0.0085 & 0.3311 & 0.4172\\
\hline
0.30 & 0.01 & -0.0038 & 0.3318 & 0.4097 & -0.0002 & 0.3331 & 0.4084 & -0.0002 & 0.3331 & 0.4084\\
0.30 & 0.02 & -0.0018 & 0.3338 & 0.4113 & -0.0071 & 0.3292 & 0.4087 & -0.0071 & 0.3292 & 0.4087\\
0.30 & 0.03 & -0.0013 & 0.3347 & 0.4127 & -0.0059 & 0.3301 & 0.4090 & -0.0060 & 0.3301 & 0.4090\\
0.30 & 0.04 & -0.0010 & 0.3353 & 0.4140 & -0.0048 & 0.3310 & 0.4093 & -0.0049 & 0.3309 & 0.4093\\
0.30 & 0.05 & -0.0009 & 0.3356 & 0.4152 & -0.0039 & 0.3316 & 0.4097 & -0.0041 & 0.3316 & 0.4097\\
0.30 & 0.10 & -0.0004 & 0.3368 & 0.4212 & -0.0021 & 0.3335 & 0.4112 & -0.0028 & 0.3331 & 0.4112\\
0.30 & 0.20 & 0.0004 & 0.3383 & 0.4334 & -0.0012 & 0.3349 & 0.4140 & -0.0044 & 0.3331 & 0.4141\\
0.30 & 0.30 & 0.0006 & 0.3397 & 0.4468 & -0.0003 & 0.3357 & 0.4165 & -0.0085 & 0.3311 & 0.4172\\
\hline
\end{tabular}
\caption{Stationary stochastic intermittent demand with $\ell=0.001,p_0=0.2$}
\label{sta4}
\end{table}

\begin{table}
\begin{tabular}{rrrrrrrrrrr}
\hline
&& \multicolumn{3}{c}{TSB} & \multicolumn{3}{c}{HES} & \multicolumn{3}{c}{LES}\\
$\alpha$ & $\beta$ & ME & MAE & RMSE & ME & MAE & RMSE & ME & MAE & RMSE\\
\hline
0.10 & 0.01 & 0.1567 & 1.4479 & 2.6820 & 0.3584 & 1.6092 & 2.7194 & 0.3579 & 1.6087 & 2.7192\\
0.10 & 0.02 & 0.0792 & 1.3929 & 2.6774 & 0.2463 & 1.5232 & 2.6975 & 0.2453 & 1.5223 & 2.6973\\
0.10 & 0.03 & 0.0480 & 1.3733 & 2.6783 & 0.1935 & 1.4843 & 2.6901 & 0.1920 & 1.4830 & 2.6898\\
0.10 & 0.04 & 0.0311 & 1.3642 & 2.6806 & 0.1613 & 1.4613 & 2.6867 & 0.1593 & 1.4595 & 2.6863\\
0.10 & 0.05 & 0.0206 & 1.3598 & 2.6834 & 0.1391 & 1.4458 & 2.6850 & 0.1366 & 1.4436 & 2.6846\\
0.10 & 0.10 & 0.0012 & 1.3577 & 2.6981 & 0.0833 & 1.4093 & 2.6848 & 0.0774 & 1.4049 & 2.6843\\
0.10 & 0.20 & -0.0040 & 1.3699 & 2.7233 & 0.0455 & 1.3888 & 2.6931 & 0.0295 & 1.3786 & 2.6926\\
0.10 & 0.30 & -0.0045 & 1.3853 & 2.7478 & 0.0336 & 1.3860 & 2.7030 & 0.0008 & 1.3668 & 2.7036\\
\hline
0.20 & 0.01 & 0.1762 & 1.4761 & 2.7141 & 0.3813 & 1.6397 & 2.7602 & 0.3809 & 1.6393 & 2.7600\\
0.20 & 0.02 & 0.0961 & 1.4186 & 2.7054 & 0.2662 & 1.5515 & 2.7323 & 0.2653 & 1.5506 & 2.7321\\
0.20 & 0.03 & 0.0627 & 1.3969 & 2.7040 & 0.2119 & 1.5113 & 2.7221 & 0.2105 & 1.5100 & 2.7218\\
0.20 & 0.04 & 0.0438 & 1.3856 & 2.7046 & 0.1787 & 1.4873 & 2.7171 & 0.1768 & 1.4856 & 2.7167\\
0.20 & 0.05 & 0.0316 & 1.3796 & 2.7062 & 0.1557 & 1.4711 & 2.7142 & 0.1532 & 1.4690 & 2.7138\\
0.20 & 0.10 & 0.0082 & 1.3741 & 2.7190 & 0.0967 & 1.4310 & 2.7104 & 0.0909 & 1.4267 & 2.7097\\
0.20 & 0.20 & 0.0031 & 1.3868 & 2.7467 & 0.0546 & 1.4063 & 2.7150 & 0.0387 & 1.3963 & 2.7139\\
0.20 & 0.30 & 0.0036 & 1.4030 & 2.7749 & 0.0407 & 1.4014 & 2.7237 & 0.0087 & 1.3831 & 2.7234\\
\hline
0.30 & 0.01 & 0.1863 & 1.4960 & 2.7504 & 0.3851 & 1.6547 & 2.8022 & 0.3847 & 1.6543 & 2.8021\\
0.30 & 0.02 & 0.1073 & 1.4394 & 2.7380 & 0.2714 & 1.5667 & 2.7697 & 0.2705 & 1.5660 & 2.7695\\
0.30 & 0.03 & 0.0735 & 1.4171 & 2.7344 & 0.2177 & 1.5267 & 2.7574 & 0.2165 & 1.5256 & 2.7571\\
0.30 & 0.04 & 0.0534 & 1.4050 & 2.7334 & 0.1850 & 1.5028 & 2.7511 & 0.1832 & 1.5013 & 2.7508\\
0.30 & 0.05 & 0.0401 & 1.3978 & 2.7337 & 0.1623 & 1.4866 & 2.7474 & 0.1600 & 1.4848 & 2.7469\\
0.30 & 0.10 & 0.0129 & 1.3885 & 2.7440 & 0.1033 & 1.4470 & 2.7407 & 0.0977 & 1.4430 & 2.7397\\
0.30 & 0.20 & 0.0066 & 1.4020 & 2.7742 & 0.0592 & 1.4204 & 2.7416 & 0.0436 & 1.4107 & 2.7399\\
0.30 & 0.30 & 0.0075 & 1.4192 & 2.8064 & 0.0440 & 1.4137 & 2.7489 & 0.0126 & 1.3958 & 2.7475\\
\hline
\end{tabular}
\caption{Decreasing demand with $\ell=0.9,p_0=0.5$}
\label{dec1}
\end{table}

\begin{table}
\begin{tabular}{rrrrrrrrrrr}
\hline
&& \multicolumn{3}{c}{TSB} & \multicolumn{3}{c}{HES} & \multicolumn{3}{c}{LES}\\
$\alpha$ & $\beta$ & ME & MAE & RMSE & ME & MAE & RMSE & ME & MAE & RMSE\\
\hline
0.10 & 0.01 & 0.0423 & 0.7610 & 1.9825 & 0.2186 & 0.9071 & 1.9962 & 0.2185 & 0.9070 & 1.9962\\
0.10 & 0.02 & 0.0160 & 0.7400 & 1.9839 & 0.1350 & 0.8401 & 1.9876 & 0.1347 & 0.8399 & 1.9875\\
0.10 & 0.03 & 0.0066 & 0.7327 & 1.9857 & 0.1045 & 0.8148 & 1.9853 & 0.1041 & 0.8144 & 1.9853\\
0.10 & 0.04 & 0.0019 & 0.7290 & 1.9874 & 0.0861 & 0.7994 & 1.9844 & 0.0855 & 0.7989 & 1.9843\\
0.10 & 0.05 & -0.0007 & 0.7270 & 1.9891 & 0.0733 & 0.7888 & 1.9840 & 0.0724 & 0.7881 & 1.9839\\
0.10 & 0.10 & -0.0046 & 0.7248 & 1.9967 & 0.0417 & 0.7634 & 1.9847 & 0.0390 & 0.7613 & 1.9844\\
0.10 & 0.20 & -0.0057 & 0.7259 & 2.0137 & 0.0229 & 0.7490 & 1.9886 & 0.0129 & 0.7415 & 1.9876\\
0.10 & 0.30 & -0.0065 & 0.7293 & 2.0342 & 0.0183 & 0.7454 & 1.9929 & -0.0054 & 0.7280 & 1.9910\\
\hline
0.20 & 0.01 & 0.0372 & 0.7557 & 1.9873 & 0.2098 & 0.9004 & 2.0027 & 0.2097 & 0.9003 & 2.0026\\
0.20 & 0.02 & 0.0123 & 0.7355 & 1.9879 & 0.1276 & 0.8334 & 1.9930 & 0.1274 & 0.8331 & 1.9929\\
0.20 & 0.03 & 0.0038 & 0.7290 & 1.9895 & 0.0979 & 0.8086 & 1.9905 & 0.0975 & 0.8082 & 1.9904\\
0.20 & 0.04 & -0.0001 & 0.7262 & 1.9912 & 0.0800 & 0.7935 & 1.9894 & 0.0793 & 0.7929 & 1.9893\\
0.20 & 0.05 & -0.0022 & 0.7251 & 1.9930 & 0.0674 & 0.7830 & 1.9888 & 0.0665 & 0.7822 & 1.9887\\
0.20 & 0.10 & -0.0040 & 0.7254 & 2.0026 & 0.0365 & 0.7578 & 1.9889 & 0.0338 & 0.7557 & 1.9885\\
0.20 & 0.20 & -0.0033 & 0.7283 & 2.0239 & 0.0187 & 0.7443 & 1.9922 & 0.0087 & 0.7368 & 1.9911\\
0.20 & 0.30 & -0.0035 & 0.7325 & 2.0475 & 0.0154 & 0.7423 & 1.9965 & -0.0081 & 0.7250 & 1.9944\\
\hline
0.30 & 0.01 & 0.0323 & 0.7520 & 1.9922 & 0.2037 & 0.8969 & 2.0101 & 0.2036 & 0.8968 & 2.0101\\
0.30 & 0.02 & 0.0074 & 0.7314 & 1.9921 & 0.1224 & 0.8301 & 1.9989 & 0.1221 & 0.8299 & 1.9989\\
0.30 & 0.03 & -0.0007 & 0.7252 & 1.9934 & 0.0930 & 0.8053 & 1.9961 & 0.0926 & 0.8049 & 1.9960\\
0.30 & 0.04 & -0.0042 & 0.7230 & 1.9951 & 0.0753 & 0.7903 & 1.9948 & 0.0746 & 0.7897 & 1.9946\\
0.30 & 0.05 & -0.0058 & 0.7225 & 1.9971 & 0.0628 & 0.7797 & 1.9940 & 0.0618 & 0.7790 & 1.9939\\
0.30 & 0.10 & -0.0058 & 0.7246 & 2.0092 & 0.0317 & 0.7539 & 1.9933 & 0.0289 & 0.7517 & 1.9929\\
0.30 & 0.20 & -0.0031 & 0.7298 & 2.0360 & 0.0134 & 0.7397 & 1.9958 & 0.0036 & 0.7324 & 1.9946\\
0.30 & 0.30 & -0.0025 & 0.7346 & 2.0635 & 0.0103 & 0.7378 & 1.9999 & -0.0127 & 0.7210 & 1.9977\\
\hline
\end{tabular}
\caption{Decreasing demand with $\ell=0.9,p_0=0.2$}
\label{dec2}
\end{table}

\begin{table}
\begin{tabular}{rrrrrrrrrrr}
\hline
&& \multicolumn{3}{c}{TSB} & \multicolumn{3}{c}{HES} & \multicolumn{3}{c}{LES}\\
$\alpha$ & $\beta$ & ME & MAE & RMSE & ME & MAE & RMSE & ME & MAE & RMSE\\
\hline
0.10 & 0.01 & 0.0426 & 0.3840 & 0.4274 & 0.0924 & 0.4174 & 0.4380 & 0.0923 & 0.4172 & 0.4380\\
0.10 & 0.02 & 0.0230 & 0.3726 & 0.4269 & 0.0637 & 0.4002 & 0.4315 & 0.0635 & 0.4000 & 0.4314\\
0.10 & 0.03 & 0.0155 & 0.3686 & 0.4278 & 0.0502 & 0.3921 & 0.4294 & 0.0499 & 0.3918 & 0.4294\\
0.10 & 0.04 & 0.0115 & 0.3668 & 0.4289 & 0.0421 & 0.3873 & 0.4286 & 0.0416 & 0.3870 & 0.4285\\
0.10 & 0.05 & 0.0090 & 0.3658 & 0.4302 & 0.0365 & 0.3841 & 0.4283 & 0.0359 & 0.3837 & 0.4282\\
0.10 & 0.10 & 0.0041 & 0.3643 & 0.4364 & 0.0229 & 0.3764 & 0.4290 & 0.0214 & 0.3755 & 0.4289\\
0.10 & 0.20 & 0.0019 & 0.3643 & 0.4492 & 0.0142 & 0.3716 & 0.4326 & 0.0097 & 0.3695 & 0.4326\\
0.10 & 0.30 & 0.0013 & 0.3642 & 0.4624 & 0.0116 & 0.3701 & 0.4367 & 0.0018 & 0.3660 & 0.4374\\
\hline
0.20 & 0.01 & 0.0426 & 0.3840 & 0.4274 & 0.0924 & 0.4174 & 0.4380 & 0.0923 & 0.4172 & 0.4380\\
0.20 & 0.02 & 0.0230 & 0.3726 & 0.4269 & 0.0637 & 0.4002 & 0.4315 & 0.0635 & 0.4000 & 0.4314\\
0.20 & 0.03 & 0.0155 & 0.3686 & 0.4278 & 0.0502 & 0.3921 & 0.4294 & 0.0499 & 0.3918 & 0.4294\\
0.20 & 0.04 & 0.0115 & 0.3668 & 0.4289 & 0.0421 & 0.3873 & 0.4286 & 0.0416 & 0.3870 & 0.4285\\
0.20 & 0.05 & 0.0090 & 0.3658 & 0.4302 & 0.0365 & 0.3841 & 0.4283 & 0.0359 & 0.3837 & 0.4282\\
0.20 & 0.10 & 0.0041 & 0.3643 & 0.4364 & 0.0229 & 0.3764 & 0.4290 & 0.0214 & 0.3755 & 0.4289\\
0.20 & 0.20 & 0.0019 & 0.3643 & 0.4492 & 0.0142 & 0.3716 & 0.4326 & 0.0097 & 0.3695 & 0.4326\\
0.20 & 0.30 & 0.0013 & 0.3642 & 0.4624 & 0.0116 & 0.3701 & 0.4367 & 0.0018 & 0.3660 & 0.4374\\
\hline
0.30 & 0.01 & 0.0426 & 0.3840 & 0.4274 & 0.0924 & 0.4174 & 0.4380 & 0.0923 & 0.4172 & 0.4380\\
0.30 & 0.02 & 0.0230 & 0.3726 & 0.4269 & 0.0637 & 0.4002 & 0.4315 & 0.0635 & 0.4000 & 0.4314\\
0.30 & 0.03 & 0.0155 & 0.3686 & 0.4278 & 0.0502 & 0.3921 & 0.4294 & 0.0499 & 0.3918 & 0.4294\\
0.30 & 0.04 & 0.0115 & 0.3668 & 0.4289 & 0.0421 & 0.3873 & 0.4286 & 0.0416 & 0.3870 & 0.4285\\
0.30 & 0.05 & 0.0090 & 0.3658 & 0.4302 & 0.0365 & 0.3841 & 0.4283 & 0.0359 & 0.3837 & 0.4282\\
0.30 & 0.10 & 0.0041 & 0.3643 & 0.4364 & 0.0229 & 0.3764 & 0.4290 & 0.0214 & 0.3755 & 0.4289\\
0.30 & 0.20 & 0.0019 & 0.3643 & 0.4492 & 0.0142 & 0.3716 & 0.4326 & 0.0097 & 0.3695 & 0.4326\\
0.30 & 0.30 & 0.0013 & 0.3642 & 0.4624 & 0.0116 & 0.3701 & 0.4367 & 0.0018 & 0.3660 & 0.4374\\
\hline
\end{tabular}
\caption{Decreasing demand with $\ell=0.001,p_0=0.5$}
\label{dec3}
\end{table}

\begin{table}
\begin{tabular}{rrrrrrrrrrr}
\hline
&& \multicolumn{3}{c}{TSB} & \multicolumn{3}{c}{HES} & \multicolumn{3}{c}{LES}\\
$\alpha$ & $\beta$ & ME & MAE & RMSE & ME & MAE & RMSE & ME & MAE & RMSE\\
\hline
0.10 & 0.01 & 0.0165 & 0.1995 & 0.3061 & 0.0661 & 0.2407 & 0.3130 & 0.0661 & 0.2407 & 0.3130\\
0.10 & 0.02 & 0.0087 & 0.1933 & 0.3067 & 0.0430 & 0.2222 & 0.3088 & 0.0429 & 0.2221 & 0.3088\\
0.10 & 0.03 & 0.0057 & 0.1910 & 0.3075 & 0.0343 & 0.2150 & 0.3076 & 0.0342 & 0.2149 & 0.3076\\
0.10 & 0.04 & 0.0042 & 0.1898 & 0.3083 & 0.0291 & 0.2105 & 0.3070 & 0.0289 & 0.2104 & 0.3070\\
0.10 & 0.05 & 0.0033 & 0.1891 & 0.3091 & 0.0254 & 0.2075 & 0.3067 & 0.0252 & 0.2073 & 0.3067\\
0.10 & 0.10 & 0.0019 & 0.1877 & 0.3131 & 0.0162 & 0.2000 & 0.3067 & 0.0155 & 0.1995 & 0.3067\\
0.10 & 0.20 & 0.0014 & 0.1871 & 0.3215 & 0.0103 & 0.1954 & 0.3078 & 0.0077 & 0.1935 & 0.3079\\
0.10 & 0.30 & 0.0012 & 0.1867 & 0.3307 & 0.0084 & 0.1939 & 0.3092 & 0.0023 & 0.1895 & 0.3095\\
\hline
0.20 & 0.01 & 0.0165 & 0.1995 & 0.3061 & 0.0661 & 0.2407 & 0.3130 & 0.0661 & 0.2407 & 0.3130\\
0.20 & 0.02 & 0.0087 & 0.1933 & 0.3067 & 0.0430 & 0.2222 & 0.3088 & 0.0429 & 0.2221 & 0.3088\\
0.20 & 0.03 & 0.0057 & 0.1910 & 0.3075 & 0.0343 & 0.2150 & 0.3076 & 0.0342 & 0.2149 & 0.3076\\
0.20 & 0.04 & 0.0042 & 0.1898 & 0.3083 & 0.0291 & 0.2105 & 0.3070 & 0.0289 & 0.2104 & 0.3070\\
0.20 & 0.05 & 0.0033 & 0.1891 & 0.3091 & 0.0254 & 0.2075 & 0.3067 & 0.0252 & 0.2073 & 0.3067\\
0.20 & 0.10 & 0.0019 & 0.1877 & 0.3131 & 0.0162 & 0.2000 & 0.3067 & 0.0155 & 0.1995 & 0.3067\\
0.20 & 0.20 & 0.0014 & 0.1871 & 0.3215 & 0.0103 & 0.1954 & 0.3078 & 0.0077 & 0.1935 & 0.3079\\
0.20 & 0.30 & 0.0012 & 0.1867 & 0.3307 & 0.0084 & 0.1939 & 0.3092 & 0.0023 & 0.1895 & 0.3095\\
\hline
0.30 & 0.01 & 0.0165 & 0.1995 & 0.3061 & 0.0661 & 0.2407 & 0.3130 & 0.0661 & 0.2407 & 0.3130\\
0.30 & 0.02 & 0.0087 & 0.1933 & 0.3067 & 0.0430 & 0.2222 & 0.3088 & 0.0429 & 0.2221 & 0.3088\\
0.30 & 0.03 & 0.0057 & 0.1910 & 0.3075 & 0.0343 & 0.2150 & 0.3076 & 0.0342 & 0.2149 & 0.3076\\
0.30 & 0.04 & 0.0042 & 0.1898 & 0.3083 & 0.0291 & 0.2105 & 0.3070 & 0.0289 & 0.2104 & 0.3070\\
0.30 & 0.05 & 0.0033 & 0.1891 & 0.3091 & 0.0254 & 0.2075 & 0.3067 & 0.0252 & 0.2073 & 0.3067\\
0.30 & 0.10 & 0.0019 & 0.1877 & 0.3131 & 0.0162 & 0.2000 & 0.3067 & 0.0155 & 0.1995 & 0.3067\\
0.30 & 0.20 & 0.0014 & 0.1871 & 0.3215 & 0.0103 & 0.1954 & 0.3078 & 0.0077 & 0.1935 & 0.3079\\
0.30 & 0.30 & 0.0012 & 0.1867 & 0.3307 & 0.0084 & 0.1939 & 0.3092 & 0.0023 & 0.1895 & 0.3095\\
\hline
\end{tabular}
\caption{Decreasing demand with $\ell=0.001,p_0=0.2$}
\label{dec4}
\end{table}

\begin{table}
\begin{tabular}{rrrrrrrrrrr}
\hline
&& \multicolumn{3}{c}{TSB} & \multicolumn{3}{c}{HES} & \multicolumn{3}{c}{LES}\\
$\alpha$ & $\beta$ & ME & MAE & RMSE & ME & MAE & RMSE & ME & MAE & RMSE\\
\hline
0.10 & 0.01 & 0.1243 & 1.3232 & 2.9427 & 0.3879 & 1.5974 & 2.9878 & 0.2161 & 1.4257 & 2.9620\\
0.10 & 0.02 & 0.0711 & 1.2661 & 2.9379 & 0.3136 & 1.5123 & 2.9680 & 0.1156 & 1.3144 & 2.9463\\
0.10 & 0.03 & 0.0509 & 1.2482 & 2.9378 & 0.2660 & 1.4616 & 2.9588 & 0.0826 & 1.2783 & 2.9416\\
0.10 & 0.04 & 0.0399 & 1.2403 & 2.9390 & 0.2325 & 1.4281 & 2.9540 & 0.0645 & 1.2603 & 2.9397\\
0.10 & 0.05 & 0.0328 & 1.2364 & 2.9408 & 0.2073 & 1.4043 & 2.9513 & 0.0526 & 1.2497 & 2.9390\\
0.10 & 0.10 & 0.0179 & 1.2337 & 2.9524 & 0.1380 & 1.3444 & 2.9494 & 0.0246 & 1.2317 & 2.9418\\
0.10 & 0.20 & 0.0120 & 1.2422 & 2.9778 & 0.0876 & 1.3075 & 2.9600 & 0.0064 & 1.2292 & 2.9552\\
0.10 & 0.30 & 0.0115 & 1.2545 & 3.0022 & 0.0698 & 1.2972 & 2.9748 & -0.0028 & 1.2322 & 2.9710\\
\hline
0.20 & 0.01 & 0.1118 & 1.3266 & 2.9865 & 0.3407 & 1.5654 & 3.0178 & 0.1898 & 1.4145 & 2.9980\\
0.20 & 0.02 & 0.0665 & 1.2777 & 2.9857 & 0.2783 & 1.4928 & 3.0061 & 0.1044 & 1.3190 & 2.9896\\
0.20 & 0.03 & 0.0487 & 1.2619 & 2.9870 & 0.2377 & 1.4491 & 3.0009 & 0.0766 & 1.2880 & 2.9878\\
0.20 & 0.04 & 0.0385 & 1.2546 & 2.9888 & 0.2085 & 1.4197 & 2.9984 & 0.0609 & 1.2723 & 2.9875\\
0.20 & 0.05 & 0.0316 & 1.2509 & 2.9908 & 0.1863 & 1.3986 & 2.9971 & 0.0503 & 1.2628 & 2.9877\\
0.20 & 0.10 & 0.0160 & 1.2473 & 3.0016 & 0.1235 & 1.3442 & 2.9972 & 0.0237 & 1.2451 & 2.9914\\
0.20 & 0.20 & 0.0088 & 1.2556 & 3.0249 & 0.0757 & 1.3101 & 3.0053 & 0.0038 & 1.2413 & 3.0015\\
0.20 & 0.30 & 0.0082 & 1.2703 & 3.0503 & 0.0586 & 1.3008 & 3.0175 & -0.0064 & 1.2435 & 3.0143\\
\hline
0.30 & 0.01 & 0.1127 & 1.3457 & 3.0353 & 0.3348 & 1.5779 & 3.0615 & 0.1879 & 1.4310 & 3.0430\\
0.30 & 0.02 & 0.0689 & 1.2983 & 3.0373 & 0.2749 & 1.5076 & 3.0540 & 0.1056 & 1.3383 & 3.0386\\
0.30 & 0.03 & 0.0515 & 1.2828 & 3.0397 & 0.2357 & 1.4652 & 3.0510 & 0.0788 & 1.3084 & 3.0387\\
0.30 & 0.04 & 0.0413 & 1.2758 & 3.0422 & 0.2073 & 1.4365 & 3.0496 & 0.0636 & 1.2929 & 3.0394\\
0.30 & 0.05 & 0.0343 & 1.2721 & 3.0446 & 0.1857 & 1.4157 & 3.0491 & 0.0533 & 1.2835 & 3.0403\\
0.30 & 0.10 & 0.0176 & 1.2679 & 3.0554 & 0.1237 & 1.3619 & 3.0505 & 0.0264 & 1.2655 & 3.0450\\
0.30 & 0.20 & 0.0087 & 1.2760 & 3.0771 & 0.0749 & 1.3256 & 3.0566 & 0.0048 & 1.2589 & 3.0528\\
0.30 & 0.30 & 0.0074 & 1.2908 & 3.1035 & 0.0566 & 1.3149 & 3.0662 & -0.0067 & 1.2593 & 3.0628\\
\hline
\end{tabular}
\caption{Sudden obsolescence with $\ell=0.9,p_0=0.5$}
\label{obs1}
\end{table}

\begin{table}
\begin{tabular}{rrrrrrrrrrr}
\hline
&& \multicolumn{3}{c}{TSB} & \multicolumn{3}{c}{HES} & \multicolumn{3}{c}{LES}\\
$\alpha$ & $\beta$ & ME & MAE & RMSE & ME & MAE & RMSE & ME & MAE & RMSE\\
\hline
0.10 & 0.01 & 0.0418 & 0.6215 & 1.7470 & 0.2120 & 0.7794 & 1.7626 & 0.1967 & 0.7642 & 1.7599\\
0.10 & 0.02 & 0.0211 & 0.6014 & 1.7483 & 0.1532 & 0.7314 & 1.7564 & 0.1092 & 0.6874 & 1.7516\\
0.10 & 0.03 & 0.0143 & 0.5953 & 1.7504 & 0.1330 & 0.7125 & 1.7542 & 0.0697 & 0.6492 & 1.7490\\
0.10 & 0.04 & 0.0110 & 0.5925 & 1.7525 & 0.1197 & 0.6993 & 1.7529 & 0.0512 & 0.6310 & 1.7480\\
0.10 & 0.05 & 0.0093 & 0.5912 & 1.7546 & 0.1093 & 0.6891 & 1.7521 & 0.0403 & 0.6203 & 1.7477\\
0.10 & 0.10 & 0.0079 & 0.5915 & 1.7634 & 0.0785 & 0.6590 & 1.7514 & 0.0189 & 0.5998 & 1.7485\\
0.10 & 0.20 & 0.0088 & 0.5958 & 1.7789 & 0.0532 & 0.6350 & 1.7547 & 0.0053 & 0.5894 & 1.7526\\
0.10 & 0.30 & 0.0088 & 0.6010 & 1.7969 & 0.0434 & 0.6256 & 1.7586 & -0.0049 & 0.5834 & 1.7564\\
\hline
0.20 & 0.01 & 0.0376 & 0.6208 & 1.7545 & 0.2054 & 0.7773 & 1.7716 & 0.1906 & 0.7625 & 1.7691\\
0.20 & 0.02 & 0.0164 & 0.6004 & 1.7551 & 0.1471 & 0.7287 & 1.7642 & 0.1043 & 0.6859 & 1.7597\\
0.20 & 0.03 & 0.0090 & 0.5939 & 1.7569 & 0.1271 & 0.7098 & 1.7618 & 0.0655 & 0.6483 & 1.7569\\
0.20 & 0.04 & 0.0055 & 0.5911 & 1.7589 & 0.1138 & 0.6968 & 1.7603 & 0.0472 & 0.6302 & 1.7558\\
0.20 & 0.05 & 0.0036 & 0.5899 & 1.7611 & 0.1035 & 0.6866 & 1.7594 & 0.0364 & 0.6197 & 1.7552\\
0.20 & 0.10 & 0.0025 & 0.5909 & 1.7718 & 0.0725 & 0.6564 & 1.7580 & 0.0144 & 0.5989 & 1.7552\\
0.20 & 0.20 & 0.0044 & 0.5962 & 1.7906 & 0.0467 & 0.6322 & 1.7605 & 0.0001 & 0.5879 & 1.7586\\
0.20 & 0.30 & 0.0048 & 0.6020 & 1.8101 & 0.0370 & 0.6233 & 1.7649 & -0.0098 & 0.5823 & 1.7631\\
\hline
0.30 & 0.01 & 0.0416 & 0.6281 & 1.7631 & 0.2236 & 0.7991 & 1.7853 & 0.2073 & 0.7829 & 1.7823\\
0.30 & 0.02 & 0.0176 & 0.6052 & 1.7631 & 0.1620 & 0.7468 & 1.7754 & 0.1150 & 0.6999 & 1.7700\\
0.30 & 0.03 & 0.0091 & 0.5978 & 1.7646 & 0.1401 & 0.7261 & 1.7722 & 0.0726 & 0.6585 & 1.7664\\
0.30 & 0.04 & 0.0048 & 0.5944 & 1.7667 & 0.1254 & 0.7116 & 1.7703 & 0.0523 & 0.6387 & 1.7648\\
0.30 & 0.05 & 0.0025 & 0.5927 & 1.7691 & 0.1139 & 0.7003 & 1.7690 & 0.0403 & 0.6269 & 1.7640\\
0.30 & 0.10 & 0.0010 & 0.5935 & 1.7824 & 0.0792 & 0.6664 & 1.7665 & 0.0156 & 0.6034 & 1.7631\\
0.30 & 0.20 & 0.0035 & 0.5995 & 1.8060 & 0.0501 & 0.6387 & 1.7682 & -0.0005 & 0.5904 & 1.7659\\
0.30 & 0.30 & 0.0040 & 0.6048 & 1.8283 & 0.0389 & 0.6285 & 1.7732 & -0.0107 & 0.5846 & 1.7712\\
\hline
\end{tabular}
\caption{Sudden obsolescence with $\ell=0.9,p_0=0.2$}
\label{obs2}
\end{table}

\begin{table}
\begin{tabular}{rrrrrrrrrrr}
\hline
&& \multicolumn{3}{c}{TSB} & \multicolumn{3}{c}{HES} & \multicolumn{3}{c}{LES}\\
$\alpha$ & $\beta$ & ME & MAE & RMSE & ME & MAE & RMSE & ME & MAE & RMSE\\
\hline
0.10 & 0.01 & 0.0470 & 0.3025 & 0.3737 & 0.1583 & 0.4195 & 0.4317 & 0.0880 & 0.3492 & 0.4008\\
0.10 & 0.02 & 0.0236 & 0.2762 & 0.3650 & 0.1242 & 0.3797 & 0.4055 & 0.0432 & 0.2987 & 0.3781\\
0.10 & 0.03 & 0.0156 & 0.2678 & 0.3629 & 0.1036 & 0.3570 & 0.3928 & 0.0286 & 0.2820 & 0.3705\\
0.10 & 0.04 & 0.0115 & 0.2638 & 0.3625 & 0.0897 & 0.3423 & 0.3857 & 0.0210 & 0.2737 & 0.3670\\
0.10 & 0.05 & 0.0090 & 0.2614 & 0.3627 & 0.0796 & 0.3319 & 0.3814 & 0.0163 & 0.2688 & 0.3651\\
0.10 & 0.10 & 0.0041 & 0.2574 & 0.3663 & 0.0526 & 0.3051 & 0.3730 & 0.0064 & 0.2593 & 0.3630\\
0.10 & 0.20 & 0.0019 & 0.2567 & 0.3768 & 0.0332 & 0.2863 & 0.3720 & 0.0008 & 0.2557 & 0.3663\\
0.10 & 0.30 & 0.0013 & 0.2565 & 0.3878 & 0.0258 & 0.2787 & 0.3751 & -0.0019 & 0.2554 & 0.3716\\
\hline
0.20 & 0.01 & 0.0470 & 0.3025 & 0.3737 & 0.1583 & 0.4195 & 0.4317 & 0.0880 & 0.3492 & 0.4008\\
0.20 & 0.02 & 0.0236 & 0.2762 & 0.3650 & 0.1242 & 0.3797 & 0.4055 & 0.0432 & 0.2987 & 0.3781\\
0.20 & 0.03 & 0.0156 & 0.2678 & 0.3629 & 0.1036 & 0.3570 & 0.3928 & 0.0286 & 0.2820 & 0.3705\\
0.20 & 0.04 & 0.0115 & 0.2638 & 0.3625 & 0.0897 & 0.3423 & 0.3857 & 0.0210 & 0.2737 & 0.3670\\
0.20 & 0.05 & 0.0090 & 0.2614 & 0.3627 & 0.0796 & 0.3319 & 0.3814 & 0.0163 & 0.2688 & 0.3651\\
0.20 & 0.10 & 0.0041 & 0.2574 & 0.3663 & 0.0526 & 0.3051 & 0.3730 & 0.0064 & 0.2593 & 0.3630\\
0.20 & 0.20 & 0.0019 & 0.2567 & 0.3768 & 0.0332 & 0.2863 & 0.3720 & 0.0008 & 0.2557 & 0.3663\\
0.20 & 0.30 & 0.0013 & 0.2565 & 0.3878 & 0.0258 & 0.2787 & 0.3751 & -0.0019 & 0.2554 & 0.3716\\
\hline
0.30 & 0.01 & 0.0470 & 0.3025 & 0.3737 & 0.1583 & 0.4195 & 0.4317 & 0.0880 & 0.3492 & 0.4008\\
0.30 & 0.02 & 0.0236 & 0.2762 & 0.3650 & 0.1242 & 0.3797 & 0.4055 & 0.0432 & 0.2987 & 0.3781\\
0.30 & 0.03 & 0.0156 & 0.2678 & 0.3629 & 0.1036 & 0.3570 & 0.3928 & 0.0286 & 0.2820 & 0.3705\\
0.30 & 0.04 & 0.0115 & 0.2638 & 0.3625 & 0.0897 & 0.3423 & 0.3857 & 0.0210 & 0.2737 & 0.3670\\
0.30 & 0.05 & 0.0090 & 0.2614 & 0.3627 & 0.0796 & 0.3319 & 0.3814 & 0.0163 & 0.2688 & 0.3651\\
0.30 & 0.10 & 0.0041 & 0.2574 & 0.3663 & 0.0526 & 0.3051 & 0.3730 & 0.0064 & 0.2593 & 0.3630\\
0.30 & 0.20 & 0.0019 & 0.2567 & 0.3768 & 0.0332 & 0.2863 & 0.3720 & 0.0008 & 0.2557 & 0.3663\\
0.30 & 0.30 & 0.0013 & 0.2565 & 0.3878 & 0.0258 & 0.2787 & 0.3751 & -0.0019 & 0.2554 & 0.3716\\
\hline
\end{tabular}
\caption{Sudden obsolescence with $\ell=0.001,p_0=0.5$}
\label{obs3}
\end{table}

\begin{table}
\begin{tabular}{rrrrrrrrrrr}
\hline
&& \multicolumn{3}{c}{TSB} & \multicolumn{3}{c}{HES} & \multicolumn{3}{c}{LES}\\
$\alpha$ & $\beta$ & ME & MAE & RMSE & ME & MAE & RMSE & ME & MAE & RMSE\\
\hline
0.10 & 0.01 & 0.0183 & 0.1781 & 0.2864 & 0.0887 & 0.2452 & 0.3057 & 0.0819 & 0.2385 & 0.3026\\
0.10 & 0.02 & 0.0090 & 0.1689 & 0.2856 & 0.0675 & 0.2270 & 0.2991 & 0.0478 & 0.2073 & 0.2934\\
0.10 & 0.03 & 0.0058 & 0.1659 & 0.2859 & 0.0590 & 0.2187 & 0.2961 & 0.0307 & 0.1905 & 0.2899\\
0.10 & 0.04 & 0.0043 & 0.1644 & 0.2864 & 0.0530 & 0.2129 & 0.2941 & 0.0225 & 0.1824 & 0.2883\\
0.10 & 0.05 & 0.0034 & 0.1635 & 0.2870 & 0.0484 & 0.2083 & 0.2927 & 0.0177 & 0.1776 & 0.2873\\
0.10 & 0.10 & 0.0019 & 0.1617 & 0.2903 & 0.0346 & 0.1945 & 0.2895 & 0.0082 & 0.1683 & 0.2861\\
0.10 & 0.20 & 0.0014 & 0.1609 & 0.2980 & 0.0229 & 0.1830 & 0.2887 & 0.0026 & 0.1634 & 0.2868\\
0.10 & 0.30 & 0.0012 & 0.1609 & 0.3070 & 0.0179 & 0.1777 & 0.2893 & -0.0010 & 0.1607 & 0.2884\\
\hline
0.20 & 0.01 & 0.0183 & 0.1781 & 0.2864 & 0.0887 & 0.2452 & 0.3057 & 0.0819 & 0.2385 & 0.3026\\
0.20 & 0.02 & 0.0090 & 0.1689 & 0.2856 & 0.0675 & 0.2270 & 0.2991 & 0.0478 & 0.2073 & 0.2934\\
0.20 & 0.03 & 0.0058 & 0.1659 & 0.2859 & 0.0590 & 0.2187 & 0.2961 & 0.0307 & 0.1905 & 0.2899\\
0.20 & 0.04 & 0.0043 & 0.1644 & 0.2864 & 0.0530 & 0.2129 & 0.2941 & 0.0225 & 0.1824 & 0.2883\\
0.20 & 0.05 & 0.0034 & 0.1635 & 0.2870 & 0.0484 & 0.2083 & 0.2927 & 0.0177 & 0.1776 & 0.2873\\
0.20 & 0.10 & 0.0019 & 0.1617 & 0.2903 & 0.0346 & 0.1945 & 0.2895 & 0.0082 & 0.1683 & 0.2861\\
0.20 & 0.20 & 0.0014 & 0.1609 & 0.2980 & 0.0229 & 0.1830 & 0.2887 & 0.0026 & 0.1634 & 0.2868\\
0.20 & 0.30 & 0.0012 & 0.1609 & 0.3070 & 0.0179 & 0.1777 & 0.2893 & -0.0010 & 0.1607 & 0.2884\\
\hline
0.30 & 0.01 & 0.0183 & 0.1781 & 0.2864 & 0.0887 & 0.2452 & 0.3057 & 0.0819 & 0.2385 & 0.3026\\
0.30 & 0.02 & 0.0090 & 0.1689 & 0.2856 & 0.0675 & 0.2270 & 0.2991 & 0.0478 & 0.2073 & 0.2934\\
0.30 & 0.03 & 0.0058 & 0.1659 & 0.2859 & 0.0590 & 0.2187 & 0.2961 & 0.0307 & 0.1905 & 0.2899\\
0.30 & 0.04 & 0.0043 & 0.1644 & 0.2864 & 0.0530 & 0.2129 & 0.2941 & 0.0225 & 0.1824 & 0.2883\\
0.30 & 0.05 & 0.0034 & 0.1635 & 0.2870 & 0.0484 & 0.2083 & 0.2927 & 0.0177 & 0.1776 & 0.2873\\
0.30 & 0.10 & 0.0019 & 0.1617 & 0.2903 & 0.0346 & 0.1945 & 0.2895 & 0.0082 & 0.1683 & 0.2861\\
0.30 & 0.20 & 0.0014 & 0.1609 & 0.2980 & 0.0229 & 0.1830 & 0.2887 & 0.0026 & 0.1634 & 0.2868\\
0.30 & 0.30 & 0.0012 & 0.1609 & 0.3070 & 0.0179 & 0.1777 & 0.2893 & -0.0010 & 0.1607 & 0.2884\\
\hline
\end{tabular}
\caption{Sudden obsolescence with $\ell=0.001,p_0=0.2$}
\label{obs4}
\end{table}


\begin{thebibliography}{10}

\bibitem{BoySyn}
Boylan, J.~E., and Syntetos, A.~A. 2007.
The Accuracy of a Modified Croston Procedure.
{\it International Journal of Production Economics\/} 107:511--517.

\bibitem{Cro}
Croston, J.~D. 1972.
Forecasting and Stock Control for Intermittent Demands.
{\it Operational Research Quarterly\/} 23(3):289--304.

\bibitem{FilEtc1}
Fildes, R., Nikolopoulos, K., Crone, S.~F., and Syntetos, A.~A. 2008.
Forecasting and Operational Research: A Review.
{\it Journal of the Operational Research Society\/} 59:1150--1172.

\bibitem{Gar}
Gardner Jr, E.~S. 2006.
Exponential Smoothing: the State of the Art --- Part II.
{\it International Journal of Forecasting\/} 22(4):637--666.

\bibitem{GhoFri}
Ghobbar, A.~A., and Friend, C.~H. 2003.
Evaluation of Forecasting Methods for Intermittent Parts Demand in the
Field of Aviation: a Predictive Model.
{\it Computers \& Operations Research\/} 30:2097--2114.

\bibitem{GooHyn}
de Gooijer, J.~D., and Hyndman, R.~J. 2005.
25 Years of IIF Time Series Forecasting: a Selective Review,
Tinbergen Institute Discussion Paper No 05-068/4, Tinbergen Institute.

\bibitem{HynKoe}
Hyndman, R. J., and Koehler, A. B. 2006.
Another Look at Measures of Forecast Accuracy.
{\it International Journal of Forecasting\/} 22(4):679--688, 2006.

\bibitem{KolSch}
Kolassa, S., and Sch\"{u}tz, W. 2007.
Advantages of the MAD/Mean Ratio Over the MAPE.
{\it Foresight: the International Journal of Applied Forecasting\/}
6:40--43.

\bibitem{LevSeg}
Lev\'{e}n, E., and Segerstedt, A. 2004.
Inventory Control With a Modified Croston Procedure and Erlang Distribution.
{\it International Journal of Production Economics\/} 90(3):361-367.

\bibitem{MakEtc1}
Makridakis, S., Andersen, A., Carbone, R., Fildes, R., Hibon, M.,
Lewandowski, R., Newton, J., Parzen, E., and Winkler, R. 1982.
The Accuracy of Extrapolation (Time Series) Methods: Results of a
Forecasting Competition.
{\it Journal of Forecasting\/} 1:111--153.

\bibitem{MakEtc2}
Makridakis, S., Chatfield, C., Hibon, M., Lawrence, M., Mills, T.,
Ord, K., and Simmons, L. F. 1993.
The M-2 Competition: a Real-Time Judgmentally Based Forecasting Study.
{\it International Journal of Forecasting\/} 9:5--23.

\bibitem{MakHib}
Makridakis, S., and Hibon, M. 2000.
The M3-Competition: Results, Conclusions and Implications.
{\it International Journal of Forecasting\/} 16(4):451--476.

\bibitem{PreEtc}
Prestwich, S.~D., Tarim, S.~A. Rossi, R., and Hnich, B. 2014.
Forecasting Intermittent Demand by Hyperbolic-Exponential Smoothing.
{\it International Journal of Forecasting\/} 30(4):928-933.

\bibitem{SnyEtc}
Snyder, R.~D., Ord, J.~K., and Beaumont, A. 2012.
Forecasting the Intermittent Demand for Slow-Moving Inventories:
A Modelling Approach.
{\it International Journal of Forecasting\/} 28:485--496.

\bibitem{Syn}
Syntetos, A. A. 2001.
Forecasting for Intermittent Demand.
Unpublished PhD thesis,
Buckinghamshire Chilterns University College,
Brunel University.

\bibitem{SynEtc3}
Syntetos, A., Babai, Z., Lengu, D., and Altay, N. 2011.
Distributional Assumptions for Parametric Forecasting of Intermittent Demand.
In: N. Altay \& A. Litteral (eds.), Service Parts Management: Demand
Forecasting and Inventory Control, Springer Verlag, NY, USA, pp.31--52.

\bibitem{SynBoy}
Syntetos, A.~A., and Boylan, J.~E. 2005.
The Accuracy of Intermittent Demand Estimates.
{\it International Journal of Forecasting\/}, 21:303--314.

\bibitem{TeuSan}
Teunter, R., and Sani, B. 2007.
On the Bias of Croston's Forecasting Method.
{\it European Journal of Operations Research\/} 194:177--183.

\bibitem{TeuEtc}
Teunter, R., Syntetos, A.~A., and Babai, M.~Z. 2011.
Intermittent Demand: Linking Forecasting to Inventory Obsolescence.
{\it European Journal of Operations Research\/} 214(3):606--615.

\bibitem{WalSeg}
Wallstr\"{o}m, P., and Segerstedt, A. 2010.
Evaluation of Forecasting Error Measurements and Techniques
for Intermittent Demand.
{\it Internation Journal of Production Economics\/} 128:625--636.

\end{thebibliography}
\end{document}